\documentclass[a4paper,11pt]{article}
\usepackage{pos,bm}

\title{Breaking the gauge symmetry in lattice gauge-invariant models}
\ShortTitle{Gauge symmetry breaking}

\author[a]{Claudio Bonati}
\author*[b]{Andrea Pelissetto}
\author[a]{Ettore Vicari}

\affiliation[a]{Physics Department, Pisa University, and INFN, sez. Pisa,\\
  L.go Pontecorvo 3, I-56127 Pisa, Italy}

\affiliation[b]{Physics Department, Sapienza University of Rome, and INFN, sez.
Roma, \\
L.go A. Moro 2, I-00185 Roma, Italy}

\emailAdd{Claudio.Bonati@unipi.it}
\emailAdd{Andrea.Pelissetto@uniroma1.it}
\emailAdd{Ettore.Vicari@unipi.it}

\abstract{We consider the role that gauge symmetry breaking terms play on
the continuum limit of gauge theories in three dimensions. 
As a paradigmatic example we consider scalar electrodynamics 
in which $N_f$ complex scalar fields interact with a U(1) gauge field. 
We discuss under which conditions a gauge-symmetry breaking term 
destabilizes the critical 
behavior (continuum limit)  of the gauge-invariant theory. We find that 
the gauge symmetry is robust at transitions at which gauge fields 
are not critical. At charged transitions, where gauge fields are critical,
gauge symmetry is lost as soon as the perturbation is added. 
}

\FullConference{%
 The 38th International Symposium on Lattice Field Theory, LATTICE2021
  26th-30th July, 2021
  Zoom/Gather@Massachusetts Institute of Technology
}

\begin{document}
\maketitle  

\section{Introduction}

Gauge symmetries play a fundamental role in the description of microscopic
phenomena, both in high-energy \cite{Weinberg-book} and 
condensed-matter physics \cite{Anderson-15,Sachdev-19}. While, in the 
first case, models enjoy an exact gauge invariance---the existence 
of an exact gauge symmetry is a basic tenet in the description of fundamental
interactions---in the second case, it may happen that the symmetry is
not an exact property of the microscopic system. It emerges at continuous 
transitions and it only characterizes the long-distance (or the low-energy
in the quantum setting) behavior of the system. 
Of course, this is possible only if 
the microscopic gauge-symmetry breaking (GSB) terms are irrelevant, in the 
renormalization-group sense, at the transition. For this reason, it is
important to understand the role that GSB terms play when added to 
gauge-invariant models. This issue is also crucial in the context of analog 
quantum simulations, when the interactions in atomic systems are engineered 
to effectively reproduce the dynamics of gauge-symmetric models, see 
Refs.~\cite{ZCR-15,BC-20} and references therein.

In this talk, we will discuss recent results
\cite{BPV-21-GSBletter,BPV-21-GSBcompact}
on the effects of GSB terms in
3D lattice gauge models with U(1) Abelian gauge invariance.
Some of the considerations presented here, however, apply also to 
non-Abelian models.

\section{Critical transitions in lattice gauge models} \label{sec2}

In this section we will briefly discuss the role that a local gauge symmetry 
plays at 3D continuous transitions.
The considerations are general and apply both to Abelian and non-Abelian
gauge models. Let $\Phi^A$ be a complex scalar field that transforms 
as 
$
\Phi^A \to \tilde{W}^{AB}(g) \Phi^B
$
under a unitary representation $W$ of a group $G$, $g$ being an element of $G$.
A $G$-invariant scalar lattice model can be defined  by the Hamiltonian (action)
\begin{equation}
H = \hbox{Re} \sum_{x\mu,A} \Phi^{A*}_{\bm x} \Phi_{{\bm x}+\hat{\mu}}^A + 
    \sum_{x} V(\Phi^2_{\bm x}), 
\end{equation}
where the first sum is over all lattice links ($\mu$ labels the lattice
directions), $\Phi^2_{\bm x} = \sum_A \Phi^{A*}_{\bm x} \Phi^{A}_{\bm x}$,
and $V(x)$ is a generic potential. By construction, this Hamiltonian is 
invariant under global $G$ transformations. 

A gauge model is obtained by selecting a subgroup $G'\subset G$ and by
associating 
group elements $U_{{\bm x},\mu}\in G'$ to each link. If 
$\tilde{U}_{{\bm x},\mu}$ 
corresponds to $U_{{\bm x},\mu}$ in the representation under which $\Phi$
transforms, we obtain the Hamiltonian
\begin{eqnarray}
H &=& \hbox{Re} \sum_{x\mu,AB} \Phi^{A*}_{\bm x} \tilde{U}_{{\bm x},\mu}^{AB}
      \Phi_{{\bm x}+\hat{\mu}}^B +
    \sum_{x} V(\Phi^2_{\bm x}) 
 + \gamma \hbox{Re} \sum_{x,\mu<\nu} \Pi_{{\bm x},\mu\nu} , 
\label{Hamgen}
\end{eqnarray}
where $\Pi_{{\bm x},\mu\nu}$ is the plaquette in the $\mu\nu$ plane built
in terms of the elements $U_{{\bm x},\mu}$. The new model is invariant under 
local transformations belonging to the group $G'$. 
 As as example, in Sec.~\ref{sec3} we will consider the compact 
Abelian-Higgs model, in which $G$ is the U($N_f$) group, 
$G'$ is the U(1) subgroup, and the fields transform under the fundamental 
representation of U(1). In this case 
$U_{{\bm x},\mu} = \exp(i \theta_{{\bm x},\mu})$, where
$\theta_{{\bm x},\mu}$  is a real number in $[0,2\pi[$
and $\tilde{U}_{{\bm x},\mu} = U_{{\bm x},\mu}$. 

Our extensive work on models with Hamiltonian (\ref{Hamgen}) shows 
\cite{PV-19-AH,BPV-19-SUN,BPV-20-SUN,BPV-20-SON,BPV-20-q2,BPV-21-noncompact,%
BFPV-21-SU}
that phase transitions occurring in gauge models can be divided 
into two broad classes.  First of all, there are transitions where 
only scalar-matter correlations are
critical. Gauge variables do not display
long-range correlations, although their presence is crucial
to identify the gauge-invariant scalar-matter critical degrees of freedom.
At these transitions, gauge fields prevent
non-gauge invariant scalar correlators from acquiring nonvanishing vacuum
expectation values and developing long-range order:
the gauge symmetry hinders some scalar degrees of
freedom---those that are not gauge invariant---from becoming critical.
The lattice Abelian-Higgs model with compact gauge
fields and unit-charge $N_f$-component
scalar fields is an example of this type of behavior~\cite{PV-19-AH}.

A second class of transitions is instead characterized by the presence 
of long-range gauge correlations. They are expected to correspond to 
the stable fixed points with nonvanishing gauge couplings (we will name 
them charged fixed points) that occur in the 
statistical field theories that are obtained in the formal 
continuum limit, i.e., that have the same field content and the same 
global and local symmetries. At present,
this type of transitions have been observed in 
Abelian models---specifically, in the lattice Abelian-Higgs model with noncompact fields 
\cite{BPV-21-noncompact} or with compact doubly-charged fields
\cite{BPV-20-q2}---and in an SU(2) gauge model with SU($N_f$) global
invariance \cite{BFPV-21-SU}.

At transitions that occur for $\gamma=0$, i.e., in the absence of the plaquette
term in Eq.~(\ref{Hamgen}), gauge fields are noncritical.
Indeed, for 
$\gamma = 0$, gauge fields can be exactly integrated out. If we define 
\begin{equation}
e^{-\beta G(\Phi_1,\Phi_2;\beta)} = 
   \int d\tilde{U} 
  \exp\left[-\beta \hbox{Re} \
   \sum_{AB} \Phi_1^{A*} \tilde{U}^{AB} \Phi_2^B \right],
\end{equation}
the model with Hamiltonian
\begin{equation}
H = \sum_{x\mu} G(\Phi_{\bm x},\Phi_{{\bm x} + \hat{\mu}};\beta) + 
     \sum_x V(\Phi^2_{{\bm x}}),
\end{equation}
is equivalent to the original one, 
as long as we consider observables that only depend on the scalar field.
Gauge invariance is still present---the function $G(\Phi_1,\Phi_2;\beta)$ 
does not vary 
if we perform gauge transformations on $\Phi_1$ and $\Phi_2$. 
This is due to the fact 
that the nearest-neighbor coupling 
$ G(\Phi_1,\Phi_2;\beta)$ can be expressed in 
terms of gauge-invariant combinations of the local fields that play the role of
order parameters. For instance, in the Abelian-Higgs U(1) case we mentioned 
above, in the London limit ($\Phi^2_{\bm x} = 1$),  
we obtain 
\begin{equation}
\int d\theta \exp\left[-\beta\hbox{Re} (\Phi_1^* e^{i\theta} \Phi_2)\right]
= I_0(\beta \sqrt{X}) \qquad
X = \sum_{AB} Q_1^{AB} Q_2^{BA} + 1/N_f,
\end{equation}
Here $I_0(x)$ is a modified Bessel function, which satisfies
$I_0(x) = 1 + x^2/4 + O(x^4)$ for small $x$, and $Q^{AB}$ is 
a gauge-invariant bilinear operator
\begin{equation}
Q^{AB}  = \Phi^{A*} \Phi^B - {1\over N_f} \delta^{AB}.
\label{Qdef}
\end{equation}
Thus, the original model is equivalent to a model with 
\begin{equation}
H_1 = - {1\over \beta} \sum_{x\mu} \ln
    I_0\left[\beta \sum_{AB} Q_{\bm x}^{AB} Q_{{\bm x} + \mu}^{BA} + 
            {\beta \over N_f} \right].
\end{equation}
Exact gauge invariance is due to the fact that the Hamiltonian only
depends on the gauge-invariant operator $Q$.
In this case the critical behavior or continuum limit is driven
by the condensation of the $Q$ operators that play the
role of fundamental fields in the Landau-Ginzburg-Wilson theory that 
should provide an effective description of the critical dynamics
\cite{PV-19-AH}. In the effective model, no gauge fields are considered.

\section{The role of GSB terms at transitions with noncritical gauge fields}
\label{sec3}

\begin{figure}[tbp]
\includegraphics[width=0.75\columnwidth, clip]{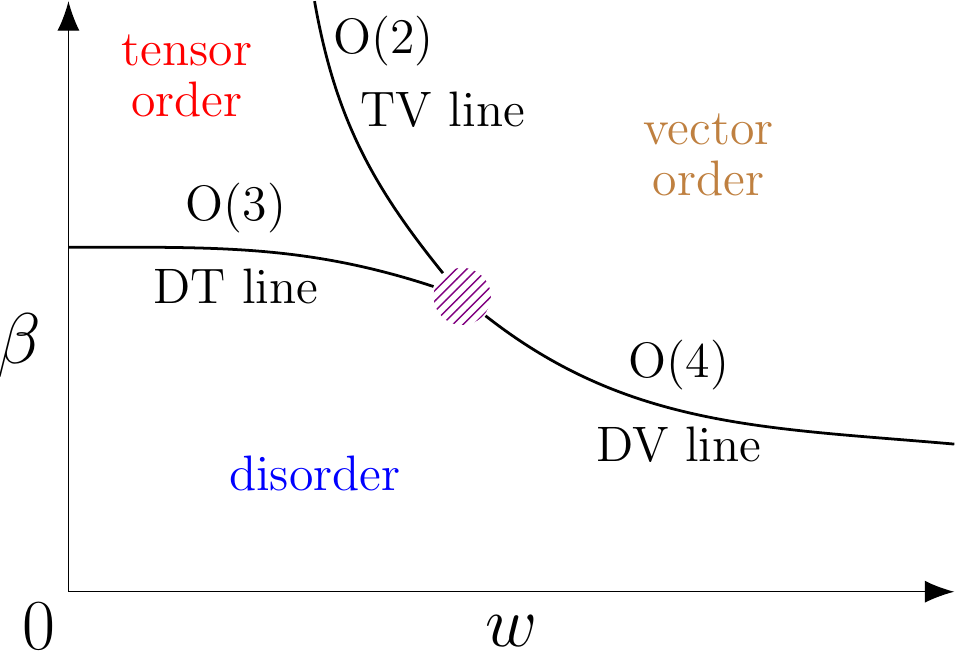}
\caption{Sketch of the phase diagram of the compact 
  lattice Abelian Higgs model 
  with $N_f = 2$, in
  the presence of the GSB term $H_{GSB}=-w\sum_{{\bm x},\mu} {\rm
    Re}\,U_{{\bm x},\mu}$ for a fixed value of $\gamma$.  
  The phase diagram is characterized by
  three different phases: a disordered phase (small $\beta$), a
  tensor-ordered phase where the tensor operator $Q$ condenses (large
  $\beta$ and small $w$), and a vector-ordered phase where the vector
  field ${\Phi}_{\bm x}$ condenses (large $\beta$ and $w$). These phases
  are separated by the disordered-tensor (DT), disordered-vector (DV), 
  and (tensor-vector) TV transition lines, where
  CP$^1$/O(3), O(4) vector, and O(2) vector critical behaviors are
  observed.}
\label{phdiagn2}
\end{figure}

Let us now consider the role played by GSB terms, considering the Abelian-Higgs
model with $N_f$ flavors in the London limit ($\Phi_{\bm x}^2 = 1$). 
As discussed in Ref.~\cite{PV-19-AH}, in this model gauge fields are never
critical: for instance,
the phase behavior is independent of the value of $\gamma$. 
For each $\gamma$, two different phases occur as $\beta$ is varied: for small 
$\beta$ there is a disordered phase, while at large $\beta$ there is an ordered 
phase, in which 
the bilinear gauge-invariant field $Q$ defined in Eq.~(\ref{Qdef})
condenses. We call this phase tensor-ordered. Because of gauge invariance,
vector correlations of the fundamental field are ultralocal, i.e.,
$\langle \Phi^{A*}_{\bm x} \Phi^{A}_{{\bm y}}\rangle = 
\delta_{{\bm x},{\bm y}}$, so that there is no vector order. 

Let us now add the GSB term 
\begin{equation}
H_{\rm GSB} = - w \sum_{{\bm x}\mu} \hbox{Re } U_{{\bm x}\mu}
\end{equation}
to the Hamiltonian. The model was studied in Ref.~\cite{BPV-21-GSBcompact} 
for $N_f=2$, obtaining the phase diagram shown in Fig.~\ref{phdiagn2}.
For small $w$ we have a low-temperature that only displays tensor order 
as for $w=0$. The corresponding order-disorder transition is the same 
as in the gauge-invariant model. Only for large values of $\beta$
does the nature of the low-temperature change. 
In this case, we have vector order, i.e., 
vector correlations of the fundamental 
field are long-ranged.  Tha analysis of Ref.~\cite{BPV-21-GSBcompact} 
shows therefore that the GSB term is 
irrelevant, in the renormalization-group sense, at the transitions occurring 
in the gauge-invariant model: the gauge-invariant behavior is robust under 
small perturbations. 

We wish now to present an argument that shows that this results is 
a general property of GSB perturbations 
at transitions where gauge fields are not
critical. Indeed, let us consider the partition function of a generic 
model with a GSB perturbation that only depends on the gauge fields:
\begin{equation}
Z = \int [dU_{{\bm x}\mu} d\Phi_{\bm x}]\,  \exp\left(-\beta H - 
    \beta H_{\rm GSB}[\{U_{{\bm x}\mu}\}] \right),
\end{equation}
where $H$ is gauge invariant.  We now perform a change of variables---therefore 
$Z$ does not change---on the scalar and gauge fields that 
corresponds to a gauge transformation. In particular, we  redefine
$U_{{\bm x}\mu} \to V_{{\bm x}} U_{{\bm x}\mu} V^\dagger_{{\bm x} + \mu}$. 
As $H$ is gauge invariant, the partition function becomes
\begin{equation}
Z = \int [dU_{{\bm x}\mu} d\Phi_{\bm x}]\, \exp\left(-\beta H -
    \beta H_{\rm GSB}[\{V_{{\bm x}} U_{{\bm x}\mu} V^\dagger_{{\bm x} + \mu} \}]
 \right).
\end{equation}
The partition function does not depend on the set of variables $V_{{\bm x}}$
and thus we can integrate over them without changing the partition function.
We define 
\begin{equation}
e^{-H_2[(\{U_{{\bm x}\mu}\}]} = 
   \int [dV_{\bm x}]\,  \exp[-
     \beta H_{\rm GSB}[\{V_{{\bm x}} U_{{\bm x}\mu} V^\dagger_{{\bm x} + \mu}
\}],
\end{equation}
and a new Hamiltonian $H' = H + H_2$. The new Hamiltonian is gauge invariant
and equivalent to the original one, if we consider the partition function and,
more generally, any
gauge-invariant correlator. The Hamiltonian $H_2$ contains interactions
between fields $U_{{\bm x},\mu}$ and $U_{{\bm y},\nu}$ at any distance 
$|x-y|$. However, for
small $\beta w$ these interactions are exponentially suppressed 
for $|x-y| \to \infty$,
and thus $H'$ represents a gauge-invariant model
with short-range interactions. To prove this crucial point, note that,
if $\beta w$ is small, one can compute $H_2$ by performing a strong-coupling
expansion. In this way, $H_2$ is written 
as a sum of lattice loops. In the expansion, a lattice loop of length $L$ is 
weighted by a factor that behaves as $(\beta w)^L$ for $\beta w \to 0$. 
For instance, the leading term is 
the plaquette, with 
a weight of order $(\beta w)^4$, which renormalizes the value of $\gamma$. 
The next term corresponds to  the $2\times 1$ plaquette, with a coefficient 
proportional to $(\beta w)^6$, and so on. Couplings therefore scale 
as $\exp[-a |x-y|]$, with $a \sim -\log (\beta w)$, proving the short-range 
nature of the interactions.

This argument proves that, for small values of $w$, 
the partition function and gauge-invariant
correlations can be computed in an equivalent gauge-invariant theory, without
GSB terms, with short-range interactions. Finally, to conclude the argument, 
let us note that we are considering a model in which gauge fields do not play
any role, i.e., the critical behavior is independent of the gauge-field
interactions: it is the same as in the original model with $\gamma = 0$, 
finally proving that the phase structure is independent of $w$. 
Note that the argument does not rely on the Abelian nature of the theory, and 
thus is should also hold in non-Abelian models.

\section{The role of GSB terms at charged transitions}
\label{sec4}

We will now discuss the role of GSB terms at charged transitions
\cite{BPV-21-GSBletter},
considering the noncompact Abelian-Higgs model with U($N_f$) global invariance.
The fundamental gauge field is a real field $A_{{\bm x}\mu}$ defined on the 
lattice links. In the London limit $\Phi^2_{\bm x} = 1$, the Hamiltonian is 
\begin{equation}
H_{nc} = \hbox{Re} \sum_{x\mu} \Phi^{*}_{\bm x}\cdot 
      \Phi_{{\bm x}+\hat{\mu}} U_{{\bm x},\mu} 
 + \gamma \sum_{x,\mu<\nu} 
    (\nabla_\mu A_{{\bm x}\nu} - \nabla_\nu A_{{\bm x}\mu})^2,
\label{Hamgen2}
\end{equation}
where ${U}_{{\bm x},\mu} = \exp(i  A_{{\bm x}\mu})$,
$\nabla_\mu f({\bm x}) = f({\bm x}+\hat\mu) - f({\bm x})$, and 
$\Phi_{{\bm x}}$ is an $N_f$-dimensional unit-length complex vector as before. 
For $N_f\ge N_f^*$, $N_f^*=7(2)$, and for a sufficiently small
gauge coupling (i.e., for $\gamma$ large enough), 
the model undergoes a transition that 
is associated with the charged fixed point of the corresponding 
field theory \cite{BPV-21-noncompact}. As expected, such a transition is not
present for small values of $\gamma$, i.e., when gauge 
fields are supposed to play no role (as we already stressed, for $\gamma=0$
they can be integrated out). 

The noncompact nature of the fields and of the gauge invariance group
(the additive group of the real numbers replaces here the compact U(1) group)
makes the discussion more complex than for the compact model. Indeed,
since the fields are unbounded, in the gauge-invariant model only gauge
invariant correlations are well-defined. Therefore, one cannot study 
the question of the relevance of the GSB perturbations directly in
the nonperturbed model. The way out of this problem is well known:
a gauge fixing should be added to make all correlations well
defined. Therefore, in the noncompact model one should consider both 
gauge-fixing terms and generic GSB perturbations.

In Ref.~\cite{BPV-21-GSBletter} we studied the effects of adding 
the perturbation 
\begin{equation}
P_M = {r\over 2} \sum_{{\bm x}\mu} A^2_{{\bm x}\mu}
\label{defPM}
\end{equation}
 to the Hamiltonian $H_{nc}$ in the presence of two different gauge fixings.
We considered the axial gauge fixing (AGF) $A_{{\bm x}3} = 0$, and 
a soft Lorentz gauge fixing (LGF), obtained by adding 
$H_{LGF}  = \sum_{{\bm x}} \exp[-a (\sum \nabla_\mu A_{{\bm x}\mu})^2]$
to the Hamiltonian.

To characterize the
strength of the perturbation $P_M$, we computed its RG
dimension $y_r>0$. This exponent provides information on how to scale
$r$ to keep GSB effects small. Indeed, when the correlation length
$\xi$ increases, approaching the continuum limit, one should decrease
$r$ faster than $\xi^{-y_r}$ to ensure that GSB effects are negligible.

A numerical finite-size scaling study shows that the 
perturbation (\ref{defPM}) is relevant at the charged fixed point
occurring for $N_f\ge N_f^*$. This is not unexpected, as this term
drastically changes 
the long-distance properties of the gauge-field correlations. 
In particular, the Coulomb phase that is present in these models 
disappears when $P_M$ is added, since its addition corresponds to adding 
a photon mass to the model.  Therefore,  as soon
as the perturbation is turned on ($r > 0$), the system flows out of
the charged Abelian-Higgs fixed point. 

However, the numerical estimates of the exponent $y_r$ 
showed an unexpected dependence on the gauge-fixing procedure.
For $N_f = 25$, we found $y_r = 2.55(5)$ for the model with
AGF, and $y_r = 1.4(1)$ for the model with LGF [for two values of 
$a$, $a=1$ and $a=10$].
The dependence of the results on the gauge fixing is puzzling and 
is presently under investigation. One possibility is that the different results 
are not due to the fact that we are considering two different gauge fixings,
the AGF and the LGF. Rather, they may be the result of the different procedure 
used.
In the axial case, field configurations
satisfy the condition $A_{{\bm x},3} = 0$, while in the Lorentz case,
the gauge-fixing term is added to the Hamiltonian, as usually done in
perturbation theory, without requiring the stronger 
condition $\sum_\mu (\nabla_\mu A_{{\bm x}\mu}) =
0$, which would correspond to $a = \infty$. Although this possibility 
might seem unlikely to perturbation-theory practitioners, the 
noncommutativity of the infinite-volume limit and of the limit 
$a\to \infty$ was already noticed in Ref.~\cite{BN-87}. They considered
the one-component Abelian Higgs model and proved that the
infinite-volume  average value of 
the scalar field---this is the expected order parameter---in the Lorentz 
gauge behaves differently for finite $a$ and for $a=\infty$.

\section{Conclusions}
\label{sec5}

In this talk we have presented our recent results on the role of GSB 
perturbations in gauge-invariant systems. At transitions in which gauge fields 
are not critical, the gauge symmetry is robust against GSB perturbations. 
If the GSB coupling is small, we still observe the same critical behavior 
as in the gauge-invariant model. In particular, the transition is still
driven by the condensation of gauge invariant observables that play the 
role of effective order parameters. 

At charged transitions (the ones where gauge fields are critical), instead,
GSB perturbations are relevant. The addition of a GSB term drives the system
out of the charged fixed point. We have studied this issue in the noncompact
Abelian Higgs model, in which gauge-dependent observables can only
be computed once a proper gauge fixing is added. The unexpected result 
is that the renormalization-group
dimension of the GSB perturbation depends on the gauge 
fixing procedure. This issue clearly requires additional work, that we hope to
present at the next-year Lattice conference.

\end{document}